\documentclass[aps,prl,twocolumn,nofootinbib,superscriptaddress,showpacs]{revtex4-1}
\usepackage{graphicx}
\usepackage{dcolumn}
\usepackage{bm}
\usepackage{color}
\usepackage{epstopdf}

\def\loga{\log_{10}(a)}

\def\etal{et al.\ }

\def\aten#1{10^{#1}}

\begin{document}


\title{Anti-Anthropic Solutions to the Cosmic Coincidence Problem}


\author{Joseph M. Fedrow}
\affiliation{Department of Astronomy, San Diego State University, San 
Diego, CA 92182} 
\affiliation{Center for Astrophysics and Space Sciences, University of 
California, San Diego, 92093}
\author{Kim Griest}
\affiliation{Department of Physics, University of California, San Diego, CA 92093, USA.}

\email{\tt j.m.fedrow@gmail.com, kgriest@ucsd.edu}


\date{\today}


\begin{abstract}

A cosmological constant fits all current dark energy data, but requires
two extreme fine tunings, both of which are currently explained by 
anthropic arguments. 
Here we discuss anti-anthropic solutions to one of these problems:
the cosmic coincidence problem- that today the dark energy density is nearly equal to
the matter density.  We replace the ensemble of Universes used in
the anthropic solution with an ensemble of tracking scalar fields that
do not require fine-tuning.  This not only does away with the coincidence problem, but also allows for a Universe that has a very different future than the one currently predicted by a cosmological constant. These models also allow for transient periods
of significant scalar field energy (SSFE) over the history of the Universe
that can give very different observational signatures as compared
with a cosmological constant, and so can be confirmed or disproved
in current and upcoming experiments.
\bigskip

\end{abstract}


\maketitle

\section{Introduction}
The nature of the dark energy (DE) currently causing the accelerated
expansion of the Universe is unknown.  A cosmological constant, $\Lambda$, 
can explain all current data, but requires two extreme fine-tunings:  the value
of $\Lambda$ must be many orders of magnitude lower than typical particle physics scales but not zero, and it must be set to just the 
right value so that the cosmic acceleration started in the recent past
as is observed \cite{carroll,padman}. 
The first fine-tuning problem above is called the cosmological constant
problem and the second fine-tuning problem is called the
cosmic coincidence problem.

The essence of the cosmic coincidence problem is that while radiation 
and matter densities drop very rapidly and at different rates as the Universe expands, a dark energy density described by a
cosmological constant stays constant throughout the entire history of the Universe.  Thus there is only one unique time
in the long history of the Universe where the DE density and matter density are roughly equal.  The cosmic coincidence is that this occurred very recently at around a
redshift of
$z\approx 0.39$. If this current epoch of cosmic acceleration had started even slightly earlier, 
the DE dominance would have stopped structure formation, and galaxies, stars,
and life on this planet would not exist.  If this epoch had been even slightly later, we would not have discovered the current accelerated expansion.

The most popular explanation for this cosmic coincidence is an anthropic
argument.  Here one imagines a large ensemble of Universes, each with
its own value of $\Lambda$.  Those Universes with values of
$\Lambda$ bigger than the current measured value do not form galaxies,
stars, life, etc. and so have no observers in them.  If the probability
for a Universe with a given value of $\Lambda$ is a strong function of 
the value of $\Lambda$, with smaller values being disfavored, then
the mostly likely Universe that has observers in it will be the Universe
with the largest value of $\Lambda$ that can form galaxies, stars, and life.
That is our Universe, so we have an explanation for the cosmic coincidence.

The general failure of non-anthropic solutions have led to this being
the most favored explanation of the cosmic coincidence problem.  
However, this solution depends upon
many unproved and perhaps unprovable hypotheses, most importantly
the existence of a huge ensemble of Universes, of which ours is
just one.

A potential problem with anthropic solutions such as this is that, if true,
it means we can never derive values of $\Lambda$ from a more fundamental
theory and perhaps many other phenomena will never be
understood from first principles.  This does not mean that the anthropic
principle cannot be correct, but
it would have been a shame if Niels Bohr had noticed that a small change
in the values of the atomic levels in atoms meant life could not exist
and had concluded that these values were therefore determined anthropically.
He might then have never discovered quantum mechanics.

Thus we are led to consider the anti-anthropic principle.
\subsection{The Anti-Anthropic Principle}
Suppose we demand that there is a non-anthropic solution to the cosmic
coincidence problem.  Is there a way to solve this fine-tuning problem
without invoking an ensemble of Universes?  As discussed by 
Griest\cite{griest02}, it is possible to replace the ensemble
of Universes with an ensemble of scalar fields which cause episodic
periods of accelerated expansion, or as discussed by Dodelson, Kaplinghat,
\& Stewart\cite{dodelson00},
with a scalar field with a complicated potential that has a similar effect.

Thus if there were many periods of cosmic acceleration, and they were
spread out over cosmic time, it would not be a coincidence that
we are currently experiencing such a period of acceleration.  In fact,
if an ensemble of scalar fields exist, each of which has a tendency
to result in a period of significant scalar field energy (SSFE), 
and these periods are spread across cosmic time, then
even if these periods of SSFE don't always give rise to accelerated
expansion, one can call this a non-anthropic solution to the cosmic
coincidence problem.

One nice feature of such an anti-anthropic solution is that it makes several
predictions, some of which may be testable:  

(i) The dark energy is not a cosmological constant.  The current period
of accelerated expansion is temporary and might finish; the DE equation
of state parameter, $w_{\phi}$, is not equal to -1, and it 
is changing with time.

(ii) There were other periods of SSFE in the past and there could be more in the
future.  These may or may not have caused periods of accelerated
expansion, but, as discussed below, these periods can still cause measurable
changes in the expansion history of the Universe.  It then becomes an experimental
question to limit or detect these periods.  

(iii) The sum of all the changing scalar field energies may eventually 
approach zero; that is, the minimum of the total potential of all these
fields may be zero, implying that the cosmological constant is actually
zero.  Thus the solution to the cosmological constant problem (as separate
from the cosmic coincidence problem) may be reduced to the older and
easier problem of finding some symmetry or reason that sets it exactly to zero.

These issues were discussed earlier by Dodelson, Kaplinghat, 
\& Steward\cite{dodelson00},
and by Griest\cite{griest02}, but the examples given by these authors
required fine-tuning. 
In particular, the toy models of Griest suffered a
severe flaw.  In order to have periods of significant scalar field energy
(SSFE), the values of the parameters in the scalar fields 
had to be finely tuned, and also the initial values of the scalar
fields themselves had to be extremely finely tuned.  Thus one solved
a fine-tuning problem by an ensemble of fine-tuned scalar fields.  Thus
the models proposed by Griest were not really a solution to the cosmic
coincidence problem.

In this paper we attempt to address and correct this flaw.  
We replace the ensemble
of monomial scalar field potentials used by Griest with an ensemble of
brane-world inspired tracking scalar fields.  These have the advantage
of having attractor-like solutions that exist independent of the initial values
of the scalar fields \cite{tracker1,tracker2}.  In addition, the form of these potentials require that the values of all parameters are of order unity in Planck units and don't need
to be finely-tuned to high precision.  The values of these
parameters do have to be set to give the current value of the dark energy density,
and to avoid conflicts with current cosmological measurements,
but this can be viewed as a measurement of the parameter values rather than a
fine-tuning.  We show some examples of such scalar field ensembles that give
SSFE over periods of interest in the Early Universe, but still agree with
current experimental measurements.  These models 
make significantly different predictions about the past and future
of the Universe than the simple cosmological constant model.

Note that several recent experimental results 
make an ensemble of scalar fields more aesthetically acceptable.
The discovery of cosmic microwave background (CMB) anisotropies  
points strongly towards an epoch of cosmic inflation in the early 
universe, most likely caused by an inflaton scalar field.  
We note that the brane-world inspired tracking potentials similar
to the ones we explore here might also make a acceptable inflaton field \cite{bento}, recent discussion of 
hilltop vs monomial potentials notwithstanding\cite{steinhardt13,linde13}.
Also, the Higgs Boson mass of around 126 GeV\cite{aad,chatrchyan} 
points to a somewhat finely-tuned scalar field sector in the Standard Model.

Thus, perhaps we should abandon our 
Occam's razor proclivities and accept that scalar fields seem to be
part of modern physics and thus may also be part of the solution 
to the DE problem.

\subsection{Experimental Constraints on Multiple Epochs of SSFE}
While the idea of many periods of accelerated expansion is appealing,
there are severe experimental constraints on such periods.
A change in the expansion history causes a change in
the relationship between distance
and redshift and also changes the growth rate of structure.  It also can change
the relative ratio of dark energy to radiation and/or matter at different
epochs.

The earliest constraint comes from Big Bang Nucleosynthesis (BBN).
By requiring that the deuterium to hydrogen ratio be within measured bounds during BBN,
Yahiro, \etal\cite{yahiro02} find that $\rho_{DE}/\rho_{rad} < 0.02$ between
$\aten{8}<z<\aten{9}$, where $\rho_{DE}$ is the energy density of DE and $\rho_{rad}$ is the energy density of radiation. Thus there cannot be a period of SSFE during
this epoch, but there can periods of SSFE before and after.

There are also strong constraints coming from the CMB.  Early constraints\cite{bean01,hu95}
have recently been updated by Linder and Smith\cite{linder11} who find 
significant changes in the CMB anisotropy power spectrum caused 
by even very short periods of accelerated
expansion.  If a period of accelerated expansion happens during, or soon after, 
recombination then peaks in the CMB power spectrum are shifted to lower
values of multipole moment, $l$, because the angular diameter distance to 
the last scattering surface decreases.  
In addition, extra decay of the gravitational potential
gives an additional Integrated Sachs-Wolfe (ISW) bump\cite{linder11}. Comparison
with the measured power spectrum rules out any period of accelerated
expansion after recombination ($z \approx 1100$).

Even a period of accelerated expansion earlier than recombination can have
important effects on the CMB power spectrum, since the sound horizon at
decoupling is decreased leading to a shift in the power spectrum peaks
to higher $l$.  In summary, Linder \& Smith find that no period of accelerated
expansion can occur after $z\approx \aten{5}$.  At higher redshifts,
they find no constraints from the CMB, so we have only the BBN constraint
above.

Accelerated expansion occurs whenever $w_{tot}<-1/3$, where $w_{tot}$ is defined in Equation 5. The limits above do not apply directly when there is SSFE which does not 
cause accelerated expansion.  However, it is beyond the scope
of this paper to calculate how much DE can exist at
various epochs with $z<\aten{5}$ without causing measurable changes
to the CMB power spectrum.  Due to the extreme precision of
recent CMB measurements, we suspect that even fairly small
amounts of scalar field DE will cause measurable changes and be excluded.
Therefore, we only consider cases where the DE has a very small fraction of the total energy density for $z < \aten{5}$.

We also must require compatibility with recent measurements of the
dark energy equation of state.  There are many results from recent
experiments, but most apply only to a $w$ whose value is constant in
time.  However, the supernova legacy survey (SNLS3) recently reported
results\cite{sullivan11}
from combining supernova measurements with WMAP7, plus the SLOAN survey
data release 7, plus
Hubble constant measurements, and found $w=-0.909\pm 0.196$ and
$w_a=-0.984\pm 1.09$, where $w(a) = w +w_a(1-a)$ is allowed to vary
linearly with the scale factor.  We can then require to around
1-sigma that our value of $w$ be less than $-0.7$, and our value of $w_a$
be between $-2.08<w_a<0.11$.

\section{Example Models}
Here we consider tracking models that don't require much 
fine-tuning.  There have been many such models suggested and we will not
review these here.  We will only consider two models from the class 
of brane-world inspired
models discussed by Dvali \& Tye\cite{dvali99}.
As a first example we consider a potential 
first discussed in detail by Albrecht and Skordis \cite{albrecht,skordis},

\begin{equation}
V(\phi)_{AS1} = V_0[(\phi - B)^2 + A]e^{-\lambda\phi} \, ,
\end{equation}

\noindent where $A, B$ and $\lambda$ are all of order unity in Planck units. We will refer to this model as AS1. In our anti-anthropic examples, because we want 
several epochs of SSFE, we introduce two more fields of identical 
form but with different values of the parameters:

\begin{equation}
V(\phi_1,\phi_2,\phi_3) = V(\phi_1)_{AS1} + V(\phi_2)_{AS1} + V(\phi_3)_{AS1} \, .
\end{equation}

Our second example is another potential studied by 
Skordis and Albrecht \cite{skordis}, 
\begin{equation}
V(\phi)_{AS2}=\left[\frac{C}{(\phi-B)^2+A}+D\right]e^{-\lambda \phi} \, ,
\end{equation}
\noindent where, as before, the parameters $A,B,C,D$ and $\lambda$ are all of order unity in Planck units. We will refer to this model as AS2. By once again 
including two additional scalar fields with potentials of identical form 
but different parameter values
we introduce additional periods of increased 
scalar field energy density,

\begin{equation}
V(\phi_1,\phi_2,\phi_3) = V(\phi_1)_{AS2} + V(\phi_2)_{AS2} + V(\phi_3)_{AS2}.
\end{equation}

\begin{figure*}[h!t!]
\centerline{
\includegraphics[width=\textwidth]{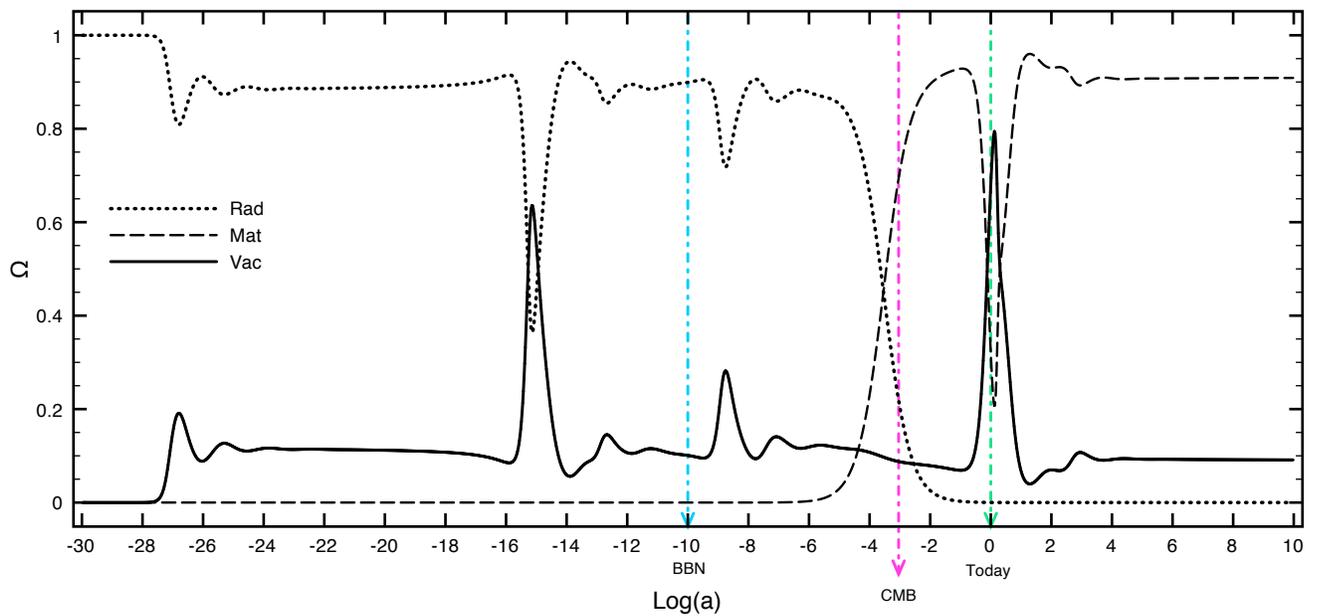}}
\caption{Transient increases in the fractional energy densities of the AS1 model with $\lambda_1=\lambda_2=\lambda_3=10$, $B_1= 14,\, B_2=16, \, B_3=27.175$, and $A_1=A_3=0.01, A_2=0.03$.}
\label{fig:as1den}
\end{figure*}

For both these example models, 
we solve the coupled Friedmann-Robertson-Walker (FRW) and scalar field equations
in the standard way, making the obvious generalization for three independent
scalar fields instead of the usual single field.
See for example, equations 1 through 7 of Skordis \& Albrecht\cite{skordis}. 
We choose parameters so that we have a transient
accelerated expansion today that gives the measured values of
Dark Energy and Dark Matter energy density, as well as a value of $w_{\phi}$ within
current limits.  We also demand that the calculated distances to
the Baryon Acoustic Oscillation (BAO) peaks agree with the measured values 
for three different redshift ranges \cite{blake12}.
In each case, we also choose parameters to give two earlier periods
of SSFE consistent with BBN and CMB constraints.

In our figures we show $\Omega_i$ vs. $\loga$, where
$a=1/(1+z)$ is the scale factor and $\Omega_i$ are the densities
of each component divided by the critical density.  We take $\Omega_{tot}=1$, consistent with observations that our Universe is flat \cite{hinshaw},
and $\Omega_{\phi}$ to be the sum of the scalar field densities
divided by the critical density. We also show $w_{tot}$ and $w_{\phi}$ vs. $\loga$, where
\begin{equation}
w_{tot}=\frac{p_r+p_m+p_{\phi}}{\rho_r+\rho_m+\rho_{\phi}} \, ,
\end{equation}
\noindent $p_i$ and $\rho_i$ are the pressure and density of radiation, matter, and the scalar fields, and $w_{\phi}=\frac{p_{\phi}}{\rho_{\phi}}$. 
The pressure of each scalar field is given by 
$ p_{\phi} = \frac{1}{2}\dot{\phi}^2 - V(\phi) \, ,$
while the density is given by
$\rho_{\phi} = \frac{1}{2}\dot{\phi}^2 + V(\phi) \,.$

Figure~\ref{fig:as1den} shows our first example.  Starting at the Planck epoch,
$\loga=-30$, corresponding to $t=10^{-43}$ s, we follow
the evolution of the radiation, matter, and scalar field energy densities by
numerically solving the differential equations that describe each.  We start 
the three scalar fields at $\phi_i=55.2$ M$_{pl}$, and ${\dot\phi_i}=0$, where the dot represents
a derivative with respect to time.  We see in Figure~\ref{fig:as1den} a few wiggles in the energy density of radiation and the vacuum soon after the Planck epoch as the tracking
behavior of the scalar field ensemble sets in. During this tracking regime $\Omega_\phi$ remains relatively constant for a long period of time.  Then, at
around $\loga \approx -16$, the first scalar field starts to dominate
and $\Omega_\phi$ rises to 0.63.  This is only a transient
epoch of SSFE and $\Omega_\phi$ drops low again as the $\phi_1$ domination
fades away.  There is another very small period of SSFE after BBN as $\phi_2$ begins to dominate and raises $\Omega_\phi$ to $0.28$ before fading away.  The third epoch of SSFE occurs as $\phi_3$ starts to dominate recently and give rise to our current period of accelerated expansion. 

In this particular model $\Omega_\phi=0.68$ today, matching the results of \textit{Planck} \cite{planck}, the Universe continues to accelerate, and  $\Omega_\phi$ reaches a maximum value of $0.79$ in the future when $a=1.33$. At this point $\Omega_\phi$ begins decreasing and at $a=2.02$ we have a second epoch of equality between $\Omega_\phi$ and $\Omega_m$. In this particular model the current epoch of accelerated expansion of the Universe will end and give rise to a secondary epoch of matter domination. This is very different from the predictions of a cosmological constant, which predicts that our current epoch of accelerated expansion will last forever. 

\begin{figure}[h!]
\begin{minipage}{\columnwidth}
\hspace{-0.5cm}
\centerline{
\includegraphics[width=\textwidth]{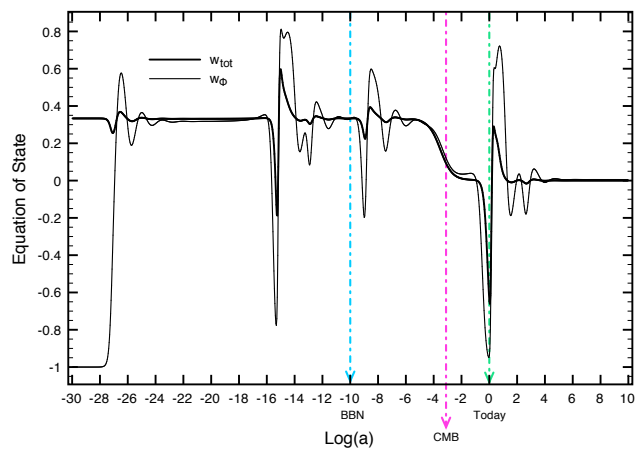}}
\end{minipage}
\caption{$w_{\phi}$ and $w_{tot}$  for the AS1 model with the same choice of parameters as shown in Figure 1.}
\label{fig:as1w}
\end{figure}

\begin{figure*}[h!t!]
\centerline{
\includegraphics[width=\textwidth]{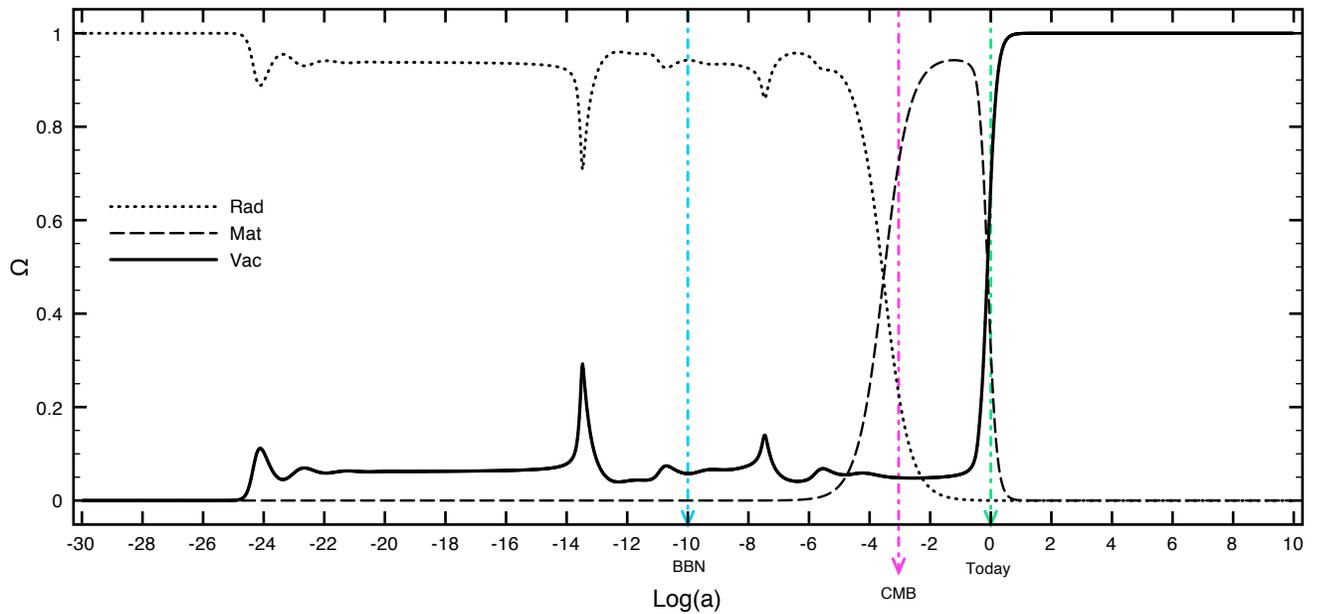}}
\caption{Transient increases in the fractional energy densities of the AS2 model with $\lambda_1=\lambda_2=\lambda_3=14$, $B_1= 12,\, B_2=16, \, B_3=20.18$, $A_1=0.004\, , \, A_2=0.01 \, ,\,  A_3=0.001 \, , \, D_1=D_2=D_3=0.1$, and $C_1=C_2=C_3=1$.}
\label{fig:AS2_density}
\end{figure*}

We can see this all from another perspective in Figure~\ref{fig:as1w} where we plot $w_i$ vs. $\loga$. We again see wiggles soon after the Planck epoch due to the onset of tracking behavior as $w_\phi$ finds its tracking solution and then both $w_{tot}$ and $w_\phi$ settle in at $1/3$ during the radiation dominated expansion.  At the same time as mentioned above ($\loga \approx -16$) $w_\phi$ changes to nearly -1 as an epoch of SSFE unfolds.  Note that $w_{tot}$ does not drop below $-1/3$ and thus there is no early period of accelerated expansion. During the second epoch of SSFE, both  $w_{\phi}$ and  $w_{tot}$ begin to drop, but, again, the Universe does not begin to accelerate. During the third epoch of SSFE $w_{\phi}$ drops to nearly -1 and $w_{tot} $ drops below $ -1/3$ giving rise to the current period of accelerated expansion.

The strength, duration, and beginning of each SSFE is set by our choice of the parameters in the AS1 potential.
For example, the $A$ parameter determines the height of each epoch of SSFE,
the $B$ parameter determines when each epoch occurs, and the $\lambda$ parameter effects the height and location of each peak. As the value of lambda becomes larger, the peaks of each SSFE become smaller and get pushed further out into the future. 

We next check whether the current accelerated expansion period predicted
by our example looks enough like a cosmological constant to satisfy 
current observations.  For this model we find today that $w=-0.94$ and
$w_a=-0.2$, within the 1-sigma constraints from SNLS3\cite{sullivan11} mentioned
above.

\begin{figure}[h!]
\begin{minipage}{\columnwidth}
\centerline{
\includegraphics[width=\textwidth]{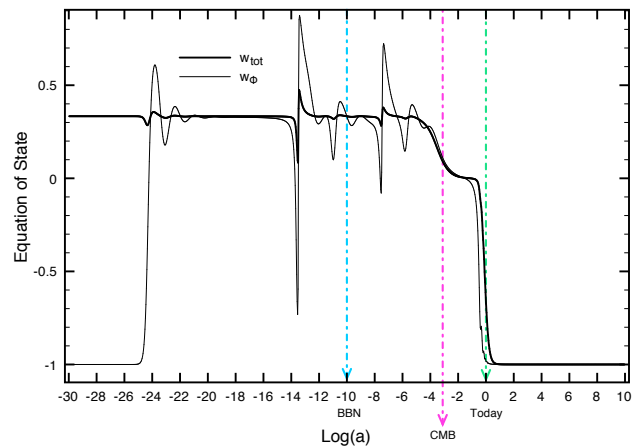}}
\end{minipage}

\caption{$w_{\phi}$ and $w_{tot}$  for the AS2 model with the same choice of parameters as shown in Figure 4.}
\label{fig:AS2_w}
\end{figure}

As our second example we calculate predictions
for a particular choice of parameters for the AS2 model.  Figure~\ref{fig:AS2_density} shows
the evolution of the $\Omega_i$ and Figure~\ref{fig:AS2_w} shows the
evolution of the $w_i$.  Even though the potential is quite different
we see fairly similar results as for AS1.  Thus one cannot say
that one potential form is greatly favored over another, and we expect
that there are other forms for the potential that would work equally
well, or better.  In Figure~\ref{fig:AS2_density} we again see wiggles in the fractional energy density of radiation and the vacuum as the tracking behavior of the vacuum sets in soon after the Planck epoch. This is followed by two periods of early
SSFE and a final period of SSFE that continues today.  In this example,
the DE dominance does continue into the future, so this set of potentials
will approach a cosmological constant model.

Figure~\ref{fig:AS2_w}
shows a similar evolution in the $w_i$ as for AS1. We find today that this model has $w_0=-0.9833$, and $w_a=0.15$.
The $w_a$ parameter is just outside the 1-sigma Sullivan, et al. \cite{sullivan11} contour, 
but this is to be expected because during the current epoch $w$ is oscillating as it drops towards $w=-1$ due to the oscillations $\phi_3$ is undergoing as it gets stuck in the potential well. 

\begin{figure}[h!]
\begin{minipage}{\columnwidth}
\centerline{
\includegraphics[width=\textwidth]{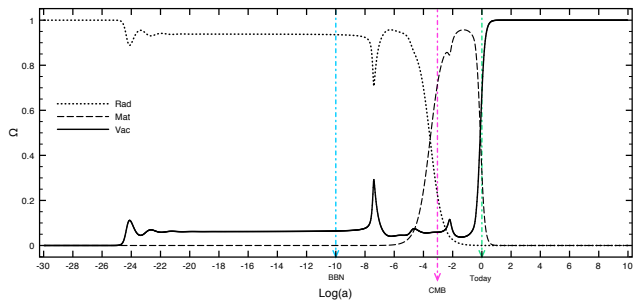}}
\end{minipage}

\caption{Transient increases in the fractional energy densities of the AS2 model with $\lambda_1=\lambda_2=\lambda_3=14$, $B_1= 16,\, B_2=19.25, \, B_3=20.18, \,A_1=0.004, \, A_2=0.01,\,  A_3=0.001, \, D_1=D_2=D_3=0.1,$ and $C_1=C_2=C_3=1$.}
\label{fig:as2primeden}
\end{figure}

\begin{figure}[ht!]
\begin{minipage}{\columnwidth}
\centerline{
\includegraphics[width=\textwidth]{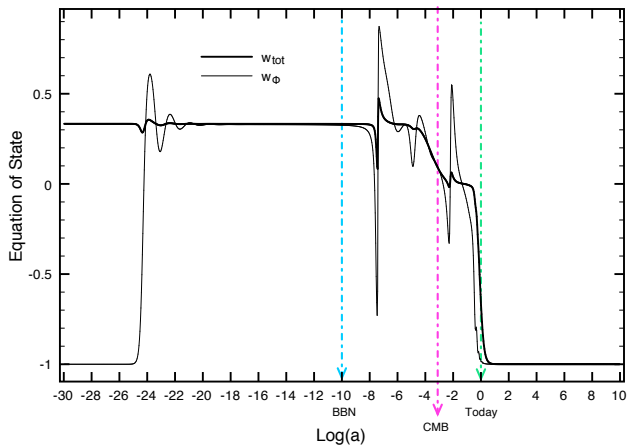}}
\end{minipage}

\caption{$w_{\phi}$ and $w_{tot}$  for the AS2 model with the same choice of parameters as shown in Figure~\ref{fig:as2primeden}.}
\label{fig:as2primew}
\end{figure}

Finally, in Figures~\ref{fig:as2primeden} and \ref{fig:as2primew} we use the AS2 potential again
and show results
for an epoch of SSFE after $\loga \approx -5$ where Linder and Smith's CMB constraints
are in full force\cite{linder11}.  The $w_i$ vs. $\loga$ plot
shows that $w_{tot}$ never gets even close to -1/3, thus there is
no period of accelerated expansion and Linder and Smith's constraint is therefore
not violated.  However, the $\Omega_i$ plot shows the density of scalar
field energy is noticeable and thus may result in
CMB, or large scale structure power spectra, that are in conflict with observations.
Addressing this question is beyond the scope of this work, but this
example shows that even without periods of accelerated expansion
one can have a cosmology that differs from a cosmological constant in
meaningful ways.

For any cosmological model to be seriously considered it must satisfy
an ever increasing set of observational measurements.  We will not attempt 
a careful check of all these recent results since we intend our models
as examples, not as proposals for the actual cosmology of the Universe.
However, we will now check our examples against the recent WiggleZ measurements
of the distance to the Baryon Acoustic Oscillation (BAO) peaks \cite{blake12}.
We view these measurements as a sort of proxy for the many 
recent experimental results.

The BAO distances can be calculated according to the formula
\cite{blake12},
\begin{equation}
\textrm{D}_{\textrm{v}} (z) = \left[\frac{z}{\textrm{H}(z)}\left(\displaystyle\int_0^{z_0}\frac{\textrm{d}z^{'}}{\textrm{H}(z^{'})}\right)^2\right]^{\frac{1}{3}} \, ,
\end{equation}
\noindent where $z$ is the redshift and $\textrm{H}(z)$ is the Hubble parameter.

\begin{figure}[ht!]
\begin{minipage}{\columnwidth}

\centerline{
\includegraphics[width=\textwidth]{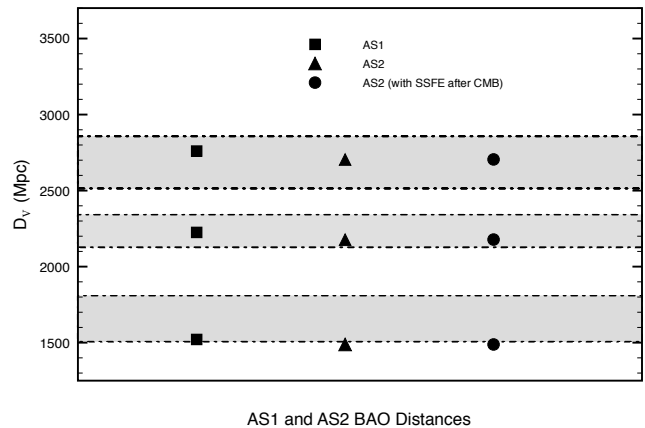}}
\end{minipage}
\caption{The calculated fiducial distances of these example models match the BAO distances observed by the WiggleZ team \cite{blake12}. The bottom grey band is the BAO peak at $z=0.44$, the middle grey band is the BAO peak at $z=0.60$, and the top grey band is the BAO peak at $z=0.73$.}
\label{fig:baolim}
\end{figure}

In Figure~\ref{fig:baolim} we show grey solid bands representing the 1-sigma
observed distances to the BAO peaks in three different redshift
intervals as measured by the WiggleZ team\cite{blake12}, as well
as our calculations of these distances in the AS1 and AS2 examples.  
We see that our calculated predicted distances match the
distances measured by the WiggleZ team.

There may be other observations that more
strongly constrain our example models, but if
we can satisfy the SNLS3 and WiggleZ constraints it is likely we can 
satisfy these
other constraints by small adjustments to our model parameters.  

\section{Conclusions}
In this paper
we solved the combined FRW/scalar field equations for two quintessence
models with multiple scalar fields designed so that they gave several
periods of significant scalar field energy (SSFE).  Such an ensemble
of scalar fields can replace the anthropic cosmological constant
model which uses an ensemble of Universes as a solution to the cosmic coincidence fine-tuning problem.  
We find that there are a wide class of
such models available including models which exhibit tracking behavior
implying that no fine-tuning of initial conditions is needed.
Such models can give different predictions from the simple anthropic
cosmological constant model and therefore
can be tested for experimentally in current and future experiments.

\begin{acknowledgments}
\acknowledgments

K.G. and J.M.F were supported in part by 
the DoE under grant DE-FG03-97ER40546.

\end{acknowledgments}

%

\end{document}